\documentclass[5p,times]{elsarticle}
\usepackage{lineno,hyperref}
\usepackage{afterpage,lipsum,capt-of,graphicx}
\usepackage{graphicx}
\usepackage{dcolumn}
\usepackage{bm}
\usepackage{hyperref}
\usepackage{float}
\usepackage{soul}
\usepackage{graphicx}
\usepackage{dcolumn}
\usepackage{bm}
\usepackage{siunitx}
\usepackage{color}
\usepackage{lipsum}
\usepackage{bm}
\usepackage{titlesec}
\usepackage{color,soul}
\usepackage{multirow}
\usepackage{siunitx}
\usepackage{color}
\usepackage{verbatim}
\usepackage{lipsum}
\usepackage[german, polish, english,slovak]{babel}
\usepackage[T1]{fontenc}
\usepackage{lipsum}  
\modulolinenumbers[5]
\usepackage{fixltx2e}
\usepackage{amsmath}
\usepackage{nidanfloat}

\journal{Arxiv}
\usepackage[colorinlistoftodos]{todonotes}

\usepackage{hyperref}
\hypersetup{
    colorlinks = true,
    linkcolor= {blue!80},
    citecolor = {red!80},
}
\usepackage[figurename=Fig.]{caption}

\addto\captionsenglish{}
\addto\captionsenglish{}
\usepackage{caption}
\usepackage{numcompress}
\usepackage{xcolor}

\begin{document}\sloppy

\begin{frontmatter}

\title{Skyrmion formation in nanodisks using magnetic force microscopy tip}

\author[Bra]{Iu. V. Vetrova}

\ead{iuliia.vetrova@savba.sk}

\author[Bra]{J. Soltys}

\author[Poz]{M. Zelent}

\author[SAR]{V. A. Gubanov}
\author[SAR]{A. V. Sadovnikov}
\author[Bra]{T. \v{S}\v{c}epka}
\author[Bra]{ J. D\'{e}rer}
\author[Bra]{V. Cambel}
\author[Bra,CEMA]{M. Mruczkiewicz}
\ead{michal.mruczkiewicz@savba.sk}

\address[Bra]{Institute of Electrical Engineering, Slovak Academy of Sciences, Dubravska Cesta 9, SK-841-04 Bratislava, Slovakia}
\address[Poz]{Faculty of Physics, Adam Mickiewicz University in Poznan, Umultowska 85, Poznan, PL-61-614 Poland}
\address[CEMA]{Centre For Advanced Materials Application CEMEA, Slovak Academy of Sciences,Dubravska Cesta 9, 845 11 Bratislava, Slovakia}
\address[SAR]{Saratov State University, Astrakhanskaya Street 83, Saratov, 410012, Russian Federation}
\begin{abstract}


In this manuscript we demonstrate the skyrmion formation in ultrathin nanodots using magnetic force microscopy tip. Submicron-size dots based on Pt/Co/Au multilayers hosting interfacial Dzyaloshiskii-Morya interaction were used in the experiments. We have found that the tip field generated by the magnetic tip significantly affects the magnetization state of the nanodots and leads to the formation of skyrmions. Micromagnetic simulations explain the evolution of the magnetic state during magnetic force microscopy scans and confirm the possibility of the skyrmion formation. The key transition in this process is the formation of the horseshoe magnetic domain. We have found that formation of skyrmion by the magnetic probe is a reliable and repetitive procedure. Our findings provide a simple solution for skyrmions formation in nanodots.


\end{abstract}


\end{frontmatter}

\definecolor{Mycolor2}{HTML}{009900}

\section{Introduction}
Magnetic skyrmions are circular domains surrounded by a single chirality domain wall~\cite{Srivastava2018Large-VoltageChirality,Je2018CreationFilms,Maccariello2018ElectricalTemperature,Heistracher2018GPU-AcceleratedAnnihilations}. They are characterized by small size and robustness against the external perturbations, which makes them attractive for modern memory-storage devices as information carriers~\cite{Ma2015,Fert_Magnetic_skyrmions_applications_2017,Suess2019SpinMemory,Suess2018AMemory}. Skyrmions can be stabilized in ultrathin films by the interplay of the external magnetic field, perpendicular anisotropy and the Dzyaloshinskii-Moriya interaction (DMI) that is induced at the interfaces of the ferromagnetic/non-magnetic metal~\cite{Romming636}. In the case of multilayer structures, there is an additional contribution from the dipolar interactions that lead to stabilization of stray field skyrmions~\cite{Buttner2018TheoryApplications,PSSR:PSSR201700259}. It was found that the confinement due to geometry can increase the stability of the skyrmion significantly~\cite{Rohart2013SkyrmionInteraction}. Therefore, low-dimensional patterned structures can serve as hosts of reconfigurable magnetic states stable at room temperature~\cite{Zeissler2017PinningStacks,Karakas2018ObservationAnisotropy}.
Measuring and controlling nucleation of skyrmions in patterned geometries still remains an important task~\cite{Zeissler2017PinningStacks, boulle2016room,Saha2019FormationAnisotropy}.

\begin{figure}[!h]
    \centering
    \includegraphics[scale=1,width=\columnwidth]{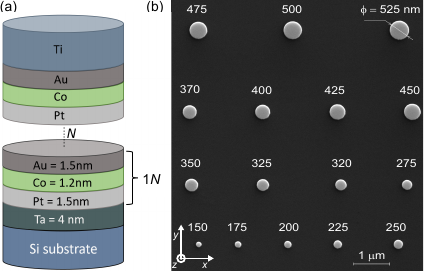}
    \caption{(a) Schematic representation of multilayer nanodot based on ultrathin Co layers placed between two different heavy metals (Au, Pt), where $N$ represents the number of trilayer Au/Co/Pt unit cell repeats. The value of $N$ is ranged from 4 to 7. (b) Scanning electron microscope top view of an array of multilayer dots patterned by the electron-beam lithography and etching method. The 16 nanodots have diameters in the range 150 -- 525 nm with the step of 25 nm. }
    \label{fig:schamatic}
\end{figure}

The challenge of the experimental study of skyrmions is to find an efficient technique for forming stable skyrmions. Recently, individual skyrmion manipulation in Pt/CoFeB/MgO multilayer using magnetic force microscopy (MFM) tip was reported~\cite{Casiraghi2019IndividualGradients}. This method was also used for nucleation of domains, skyrmions or skyrmion lattices in various non-patterned ultrathin films  \cite{Qin2018Size-tunableMultilayers, Temiryazev2018FormationMicroscope, Zhang2018DirectField}. The effect of the reversal magnetization of skyrmions by MFM tip was studied in symmetric Pt/Co/Pt thin films consisting of modified cylindrical regions with a diameter of 100 nm~\cite{Mironov2019ImpactAnisotropy}. Highly metastable hedgehog skyrmions in soft magnetic permalloy nanodots with a diameter of 70 nm and height of 30 nm was presented in the Ref.~\cite{Berganza2018ObservationNanodots}.

Here we show for the first time that the MFM tip can be used to form the skyrmions in the nanodots patterned from multilayer ultrathin films. Skyrmions are induced in our structures with high yield. We elucidate the influence of the dot diameter and the number of repetition of the layers on the magnetic state before and after nucleation realized by the magnetic probe. Then, with numerical simulations, we explain the process of the skyrmions formation.

\section{Methods}
\subsection{\textit{Micromagnetic simulations}}
The skyrmion formation processes in a nanodots were simulated using Mumax3~\cite{MuMax2011_main,mumax_2014,Leliaert2014}. The simulations were performed for disk diameter ranged from 100 to 400 nm using a uniformly discretized grid with a cell size of 1.0 nm $\times$ 1.0 nm $\times$ (1.0--1.3)~nm. For the simulations we employed following parameters: exchange constant $A=1.5 \times 10^{-11}$ J/m, interfacial Dzyaloshinskii-Moriya interaction $D= 1.0$ mJ/$\textrm{m}^{2}$, magnetization saturation $M_\mathrm{s}=1.2 \times 10^{6}$~J/$\mathrm{m^{3}}$, and a perpendicular magnetic anisotropy of $K_\mathrm{u}=3.45 \times 10^{5}$~J/$\mathrm{m^{3}}$. In order to mimic the polycrystalline nature of the sample, a Voronoi tessellation was added in the micromagnetic simulations in which to each 6 nm grain a slightly different perpendicular magnetic anisotropy was assigned. The anisotropy value was choosen randomly from a normal distribution centered about a mean value with a standard deviation of $2.5\%$ \cite{Zeissler2017PinningStacks,Saha2019FormationAnisotropy}. To simulate the skyrmion nucleation process in nanodots, we exclude exchange interactions between the ferromagnetic layers. In order to provide insight into the skyrmion formation in nanodots, micromagnetic simulations were performed for different thicknesses of ferromagnetic layers in the range from 1.1 to 1.3 nm. 

\subsection{\textit{Fabrication}}

The experimental samples with multilayer nanodots are based on dipolary coupled ultrathin Co layers with four, five, six and seven repetitions of the Pt/Co/Au tri-layer (see Fig. 1 (a)). The Pt/Co/Au-based tri-layer was chosen so that both dipole and DMI interactions influence the static micromagnetic state. All layers were sputtered in-situ on a Si substrate by DC Magnetron Sputtering at room temperature (base pressure $\sim 8 \times 10^{-5}$ Pa; working gas - Ar pressure $\sim 0.175$ Pa). The thicknesses of the deposited layers ($\mathrm{t}_{\mathrm{Pt}}=$1.5 nm, $\mathrm{t}_{\mathrm{Co}}$=1.2 nm, $\mathrm{t}_{\mathrm{Au}}$=1.5 nm) was defined by the rotation speed of the substrate. Then electron beam lithography with a positive tone resist was used to define multiple arrays of dots with variable diameters varied from 150 to 525 nm. After the development, a 15 nm-thick titanium layer was evaporated and lifted-off to reveal Ti circle-shaped mask patterns. Finally, Kaufmann-type Ar-ion beam etching was used to transfer the Ti-mask pattern into the Pt/Co/Au multilayer resulted in the formation of stacked nanodots. The total height of the nanodots was 45, 52, 58 and 64 nm for samples with  4, 5, 6 and 7 repetitions of the Pt/Co/Au tri-layer, respectively (see Fig. 1 (b)).

\subsection{\textit{Measurements}}
Magnetic characterization of the representative sample was carried out using a vibrating sample magnetometer with an in-plane magnetic field up to 2 T. The obtained value of the magnetization saturation was about $\mathrm{M_{s}}=1.3$~MA/m, which is slightly lower than the bulk value of the $\mathrm{M_{s}}$ for cobalt (1.4~MA/m)~\cite{Bautin2017MagneticNanoparticles}. We expect that the $\mathrm{M_{s}}$ does not strongly depend on the number of trilayer repetition. The measurement for 7 repetitions was performed and this value of $\mathrm{M_{s}}$ was used to calculate DMI.
\hfill \break

The DMI strength was evaluated by the Brillouin Light Scattering (BLS) measurements in the Damon-Eshbach configuration~\cite{bls1,bls2,bls3}. By this system, one can measure the frequency shift $ \Delta f_{\textrm{SW}}$ between the counter-propagating spin waves at the specific value of the wavenumber  $k_{\textrm{SW}}$.The value of $\Delta f_{\textrm{SW}}$  corresponds to the shift between the Stokes and anti-Stokes frequencies, and then the value of DMI constant $D$ can be directly estimated as:
\begin{equation}
  D=\frac{\Delta f_{\textrm{SW}} \pi \mathrm{M_s}} {2\gamma k_{\textrm{SW}}},
    \label{eq:1} 
\end{equation}
where $\gamma=176$~GHz/T is the gyromagnetic ratio.
The measured values of frequency shift and  DMI constant for the spin waves with the wavevector  $k_{\textrm{SW}}=0.011 \mathrm{nm}^{-1}$ was calculated using the Eq.~\ref{eq:1} and are summarized in Table 1. 
\begin{table}
\begin{center}
\caption{The measured value of frequency shift and calculated DMI constant}
\label{Table1}
\begin{tabular}{|c|c|c|} \hline
Sample            & $\Delta f_{\textrm{SW}},~\textrm{GHz}$& $D,~\mathrm{mJ/m}^{-2}$   \\ \hline
(Pt(1.5)/Co(1.2)/Au(1.5))$_{x4}$ & 1.12 $\pm \textrm{\space}$ 0.1        & 1.19 $\pm  \textrm{\space}0.1 $    \\ 
(Pt(1.5)/Co(1.2)/Au(1.5))$_{x5}$ & 1.16 $\pm  \textrm{\space}0.1$       & 1.23 $\pm  \textrm{\space}0.1 $    \\ 
(Pt(1.5)/Co(1.2)/Au(1.5))$_{x6}$ & 0.89 $\pm  \textrm{\space}0.1$       & 0.95 $\pm  \textrm{\space}0.1  $   \\ 
(Pt(1.5)/Co(1.2)/Au(1.5))$_{x7}$ & 1.12$ \pm  \textrm{\space}0.1$       & 1.19 $\pm  \textrm{\space}0.1 $  \\ \hline
\end{tabular}
\end{center}
\end{table}
The magnetic states of multilayered nanodots were investigated and induced by MFM with low magnetic moment tip (LM) and high magnetic moment tip (HM) respectively. The measurements were performed at room temperature with the tip scanning speed of 5 um/sec. In both LM and HM scans, a standard two-pass MFM method was used where the tip scanned over each line four times as follows: At first, sample topography was recorded in semi-contact mode (1st pass). During this scan MFM tip was moving along the $x$-axis: from left to the right side of a dot array, then it was returned to the beginning of the same line. Within the second pass, the tip was lifted (in $z$-direction) at a distance of 15 nm above the sample to lower the Van der Waals forces and to image only magnetic forces. Next, the tip position was shifted down by 15 nm (along $y$-axis) to continue scanning on the next line.  

All samples were firstly scanned by LM super-sharp silicon probes (SSS-MFMR, $Nanosensors^{TM}$) with tip radius below 15 nm optimized for high-resolution imaging ($\sim$ 20 nm). The magnetic coating of these probes is characterized by a very low effective magnetic moment $ 80 \times 10^{3} $ A/m, which helps to minimize the invasive interaction between the tip and sample. The scans taken by LM probes revealed initial magnetic states in as-fabricated dots without affecting it by measurements. Subsequently, the same arrays were scanned by the HM probe with approximately four times higher magnetic moment (MESP, Bruker). Before scanning HM probes were magnetized in the external magnetic field in order to set them into a single domain state. Finally, remagnetized magnetic states were measured by LM probe again. Four identical dot arrays on each sample were scanned in order to confirm the reproducibility of skyrmion formation.

\begin{figure}[tp!]
    \centering
    \includegraphics[scale=1,width=\columnwidth]{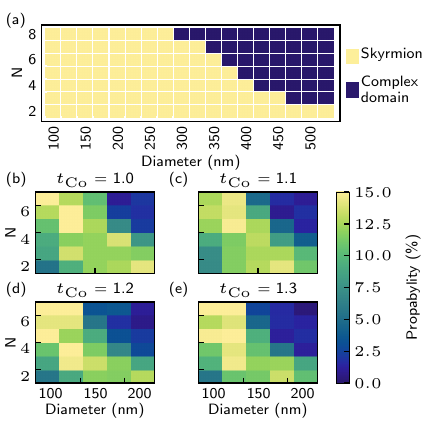}
    \caption{Simulation of skyrmion stability in nanodots: a) Skyrmion stability range. b) An estimated probability of skyrmion formation in nanodots in dependence on ferromagnetic layer thickness, a number of repetition, and nanodot diameter. }
    \label{fig:statistics}
\end{figure}
\begin{figure*}[!bp]
    \centering
    \includegraphics[scale=1,width=0.95\textwidth]{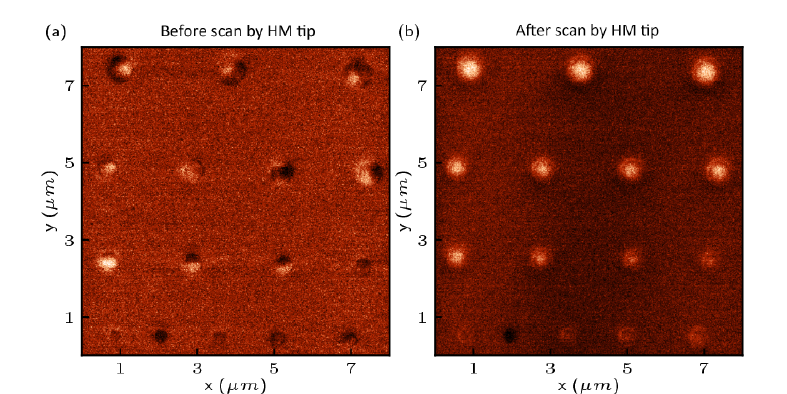}
    \caption{Results of the MFM measurements with LM tip in array of nanodisks (a) before and (b) after scan by HM tip. A complex domain structures are observed in (a), including horseshoe-like domains (i.e. 525, 500, 475, 375 nm). Magnetic skyrmions are observed in all disks above diameter  200 nm after the HM scan (b).}
    \label{fig:mfm}
\end{figure*}

\section{Results and discussion}

Firstly, we verified the range of skyrmion stability depending on the number of repetition, for $\mathrm{N} = \langle 2,8 \rangle$ and the diameter of the nanodisk in range $\mathrm{D} = \langle 100, 525 \rangle$ nm with numerical simulations. It was shown that skyrmion is a stable state in nanodots over a wide range of geometric parameters. It was observed that with the increase of the number of multilayers and dot diameter, i.e. with the increase of the strength of dipole interactions, skyrmions are not stable (see Fig. ~\ref{fig:statistics} (a)). We expect that the range of skyrmion stability in experimental measurements may vary, due to sample nonuniformities, polycrystalline structure or assumptions included in micromagnetic simulation~\cite{Zeissler2018DiscreteNanodiscs}. Nevertheless, Fig. ~\ref{fig:statistics} (a) demonstrates an approximate boundary of skyrmion stability in samples of our interest.

To investigate the probability of skyrmion spontaneous formation in dependence on the geometrical parameter for each combination of diameter and number of repetition, we performed relaxation procedure for 500 different random initial states, where each grain inside ferromagnetic layer has a different orientation of the magnetization. For each case, we started simulations with the same initial pseudo-random number, which determine the statistical distribution of magnetization in the grains. In these simulations, we neglected an antisymmetric exchange interaction to investigate the probability under unfavourable skyrmions conditions. Simulations, including these interactions, did not show a significant increase of  the skyrmion nucleation probability. Micromagnetic simulations show that the highest probability state for each case is a complex domain state, with a low probability for the single-domain state.
Furthermore, we have found that the diameter and number of repetition has a significant influence on skyrmion formation probability. The domain structure was observed mostly in the dots with diameters larger than 200 nm, the single skyrmion with probability less than 15\% states was found only for a few sets of parameters, where the diameter was less than 200 nm (see Fig.~\ref{fig:statistics} (b-e)). Our numerical calculations agree with experimental MFM measurements performed with LM tip in all samples. An example scan of one array of the as-fabricated sample (without a history of the external magnetic field) with six repetitions is shown in Fig. \ref{fig:mfm} (a), where only complex domain states have been found. 

\begin{figure}
    \centering
    \includegraphics[scale=1,width=\columnwidth]{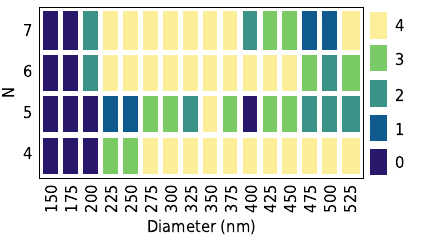}
    \caption{The total number of observed skyrmions induced by high momentum tip on four arrays per sample.}
    \label{fig:N_VS_Dd}
\end{figure}

Next, we compare the results of LM MFM measurements of magnetic states before and after HM tip scan (Fig. \ref{fig:mfm} (a) and (b) respectively). Both LM measurements and HM tip scan was performed with two pass techniques as described in the methods section. We have observed that the magnetic states were strongly influenced by HM tip scan. Skyrmions have been induced in disks with a radius larger than 200 nm. The same procedure (LM measurements, HM scan and LM measurements) was repeated for four identical arrays with disk diameter between 150-525 nm for four different samples. Fig ~\ref{fig:N_VS_Dd} presents the total number of observed skyrmions induced by high momentum tip on four arrays per sample. The number of counts of observed skyrmions is increasing for small dot diameters (200 nm) as the number of repetition is increasing, whereas it is decreasing for large dot diameters (> 450 nm). Experimental observation can be related to lower stability of skyrmions for the high number of the repetitions of layers and high diameters, that is observed in Fig.~\ref{fig:statistics} (a). From the Fig ~\ref{fig:N_VS_Dd}, we can also determine the critical dot size (175 nm) below which skyrmions were not observed in any samples. A possible reason for this is that for small disks the MFM tip acts as uniform field over a relatively small area of the dot, not locally. As it will be explained in the next part of manuscript, local character of the field is crucial for the skyrmion nucleation.

We perform numerical simulations to understand the mechanism of the isolated skyrmion formation during the HM tip scan procedure. The influence of tip field induced by the MFM tip and pinning at the dot boundary have to be considered. Due to the lack of exact experimental data about the polarity, spatial distribution, and magnetic field strength of the MFM tip, we performed a series of simulations to check different sets of parameters. We approximate the magnetic field profile induced from the tip as a Gaussian function (homogeneous through the thickness of the sample with 25 nm full width at half maximum). In the experiment the MFM tip was not perpendicular to the sample, its axis was tilted at 75 deg with respect to the sample surface, and we adapted this tilting also into simulation. In the next part of our manuscript, the magnetic field induced by the tip will be called the tip field. 

In order to take into account the pinning effects at the dot edges, we have used the modified code of the Mumax3. We defined outer ring of 10 nm width. In this region we add a Voronoi tessellation where in each 6 nm grain a different value of the perpendicular magnetic anisotropy was assigned. The value of anisotropy in this grains was chosen with a normal distribution centered about a nominal value with a standard deviation of $30\%$.

The origin of the boundary pinning might be due to nonuniformities at boundaries ~\cite{Bryan2005EdgeNanostructures,Putter2009TheElements}, thickness nonuniformities~\cite{Maranville2007VariationStripes} or oxidation's at the edges~\cite{Zhu2010ModificationEdges,Guo2013SpectroscopyNanodisks}.

In the simulations, we have reproduced the MFM scan path by scanning each horizontal line twice, from left to right, right to left, and then moving to the next line. The horizontal and vertical step was set to 15 nm. We have performed simulations for various tip fields 100, 200 and 300 mT. During the experiment the MFM tip collects the data from each location for few seconds. To reproduce this procedure in simulations, we relaxed magnetization for each position of the tip field.

\begin{figure*}
    \centering
    \includegraphics[scale=1,width=\textwidth]{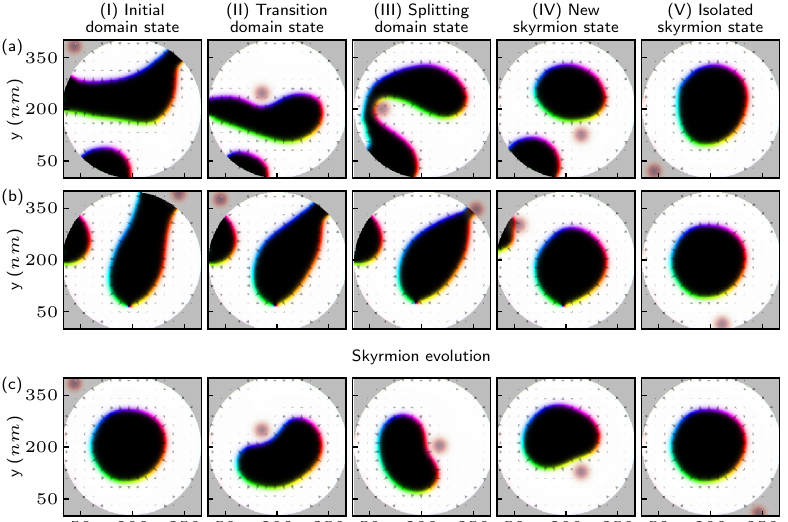}
    \caption{Simulated images of the evolution of the magnetic configuration in nanodots in dependence on the initial state (I) and different positions of the tip field. Simulations were performed for 400 nm disks with 4 repetition of the stack with DMI=1.0 $mJ/m^2$. The red circle represents the z-component of the magnetic field induced by the high-momentum tip. The amplitude of tip field was 200 mT at center of the spot, and spot diameter was equal to 25 nm. The colors represent the orientation of the magnetization.} 
    \label{fig:three_cases}
\end{figure*}

We have observed that the tip field strength of 100 mT is not sufficient to re-magnetize or move the domain. The tip field with strength 200 mT is strong enough to move and annihilate domains at the boundaries. However, it is not sufficient to remagneitze the large domain or SD state with opposite polarity The field  300 mT is sufficient to re-magnetize domain with any size (also SD state). The domain behaviour at these selected fields observed in the simulations may differ in relation to experimental measurements, taking into account magnetic material parameters or heterogeneity of anisotropy~\cite{Metaxas2007,Mironov2009MagneticAnisotropy,Gorchon2014Pinning-dependentFilm}.

In the case when the value of the tip field was over 200 mT, the MFM scan nucleates skyrmions. Exemplary evolution of the magnetization in 400 nm disks during the HM scan is shown in Fig.~\ref{fig:three_cases} for three different initial states at 200 mT. The polarization of the tip field in these simulations corresponds to the polarization of domain with positive polarization (marked with white color). Therefore the tip field will lead to expansion of the domains with positive polarization.  In Fig. 5 (a, I-II), it is shown how the tip pushes the domain with negative polarization (black domain) down the nanodot as the scanning proceed from top to bottom. This leads to an intermediate state with large domain with positive polarization filling the upper half of the disk and mixed domain pattern confined to lower part of the disk, see the Fig. 5. (a, II). When the tip field moves away from the sample, the domain with negative polarity fills the empty space in the middle of the disk, which forms the horseshoe state (see the Fig. 5 (a, III)). Next, the magnetic tip continuous the scan in lower line and splits the horseshoe domain to a skyrmion and a remaining domain, see the Fig. 5 (a, IV). The result is skyrmion with opposite polarity to the magnetic tip. In next steps, the tip field annihilates the remaining domains as the scan continue, see the Fig. 5 (a, V). 

An analogous process of domain displacement relative to the local field was observed regardless of the initial state. In all simulations where we assumed edge pinning, we did not observe domains near the edges in the final configuration, which is in good agreement with the experimental results. 
The process  of skyrmion splitting from horseshoes state and annihilation remaining domains has been observed when the domain pinning at the edge of the nanodot was significant. In samples without pinning we observed rotation of horseshoe state instead of splitting of skyrmions. 

We also studied numerically whether the scanning HM-MFM can destroy skyrmion. We performed simulations, assuming Bloch type skyrmion as the initial state (see Fig. ~\ref{fig:three_cases} (c)). As the field moved, the skyrmion was pushed from its central position, contracting or stretching around the tip field, see Fig. 5. (c). These results shows that repeated HM-momentum tip scanning can lead to an increase in skyrmion formation probability.



\section{Conclusion}

We have been investigated a skyrmion formation in  Pt/Co/Au nanodots. We demonstrate that the high magnetic momentum probe induces individual skyrmions during the scan. A specific path of the MFM tip causes a systematic change in the state of the magnetic domains during scanning. Regardless of the initial domain state, the tip field induced by the probe leads to the formation of transient, horseshoe state. The skyrmion is split out from the horseshoe state. The process of annihilation of remaining domains is ruled by magnetostatic interactions where the magnetic domains are simultaneously repelled from the edges of the nanodots and from the tip field.  We observed that the inhomogeneous of the distribution of anisotropy reduces the possibility of domain movement and leads to faster annihilation. This increases the chances of the central domain being seen as an isolated skyrmion.  Our proof-of-concept results pave the way towards fast and effective individual skyrmion nucleation in nanodots.  This method is primarily aimed as a customization tool for preparing and studying the limited number of skyrmions due to a slow-scanning speed of the tip. However, simultaneous writing/reading process with multiple arrays of tips can increase throughput significantly~\cite{Vettiger2002TheStorage}. This could lead to the development of a new concept of the memory device based on magnetic scanning probes that combine ultrahigh density and high-speed data transfer rate.

\section*{Acknowledgement}
This work was financially supported by APVV project no. APVV-16-0068 (NanoSky), Vega project 2/0160/19, Era.Net RUS Plus (TSMFA), RFBR (projects no. 18-29-27026, 18-57-76001) and Grant of the President of RF (MK-1870.2020.9,SP-2819.2018.5). BLS measurement of counter-propagating spin waves was performed in the framework of  RSF(project no. 16-19-10283). M.Z. acknowledge funding from the National Science Centre of Poland Grant No. 2017/27/N/ST3/00419, 2018/30/Q/ST3/00416, Adam Mickiewicz University Foundation. The simulations were partially performed at the Poznan Supercomputing and Networking Center (Grant No.~398). This study was performed in the frame the implementation of the project Building-up Centre for advanced materials application of the Slovak Academy of Sciences, ITMS project code 313021T081 supported by Research \& Innovation Operational Programme funded by the ERDF.


\section*{References}
\begin{thebibliography}{39}
\providecommand{\natexlab}[1]{#1}
\providecommand{\url}[1]{\texttt{#1}}
\providecommand{\href}[2]{#2}
\providecommand{\path}[1]{#1}
\providecommand{\eprint}[1]{\href{http://arxiv.org/abs/#1}{\path{#1}}}
\providecommand{\DOIprefix}{doi:}
\providecommand{\ArXivprefix}{arXiv:}
\providecommand{\URLprefix}{URL: }
\providecommand{\Pubmedprefix}{pmid:}
\providecommand{\doi}[1]{\href{http://dx.doi.org/#1}{\path{#1}}}
\providecommand{\Pubmed}[1]{\href{pmid:#1}{\path{#1}}}
\providecommand{\BIBand}{and}
\providecommand{\bibinfo}[2]{#2}
\ifx\xfnm\undefined \def\xfnm[#1]{\unskip,\space#1}\fi
\bibitem[{Srivastava et~al.(2018)Srivastava, Schott, Juge,
  K{\v{r}}i{\v{z}}{\'{a}}kov{\'{a}}, Belmeguenai, Roussign{\'{e}}
  et~al.}]{Srivastava2018Large-VoltageChirality}
\bibinfo{author}{Srivastava\xfnm[ T.]}, \bibinfo{author}{Schott\xfnm[ M.]},
  \bibinfo{author}{Juge\xfnm[ R.]},
  \bibinfo{author}{K{\v{r}}i{\v{z}}{\'{a}}kov{\'{a}}\xfnm[ V.]},
  \bibinfo{author}{Belmeguenai\xfnm[ M.]},
  \bibinfo{author}{Roussign{\'{e}}\xfnm[ Y.]}, et~al.
\newblock \bibinfo{title}{{Large-Voltage Tuning of Dzyaloshinskii-Moriya
  Interactions: A Route toward Dynamic Control of Skyrmion Chirality}}.
\newblock \bibinfo{journal}{Nano Letters}
  \bibinfo{year}{2018};\bibinfo{volume}{18}(\bibinfo{number}{8}):\bibinfo{pages}{4871--4877}.
\newblock \DOIprefix\doi{10.1021/acs.nanolett.8b01502}.
\bibitem[{Je et~al.(2018)Je, Vallobra, Srivastava, Rojas-S{\'{a}}nchez, Pham,
  Hehn et~al.}]{Je2018CreationFilms}
\bibinfo{author}{Je\xfnm[ S.G.]}, \bibinfo{author}{Vallobra\xfnm[ P.]},
  \bibinfo{author}{Srivastava\xfnm[ T.]},
  \bibinfo{author}{Rojas-S{\'{a}}nchez\xfnm[ J.C.]},
  \bibinfo{author}{Pham\xfnm[ T.H.]}, \bibinfo{author}{Hehn\xfnm[ M.]}, et~al.
\newblock \bibinfo{title}{{Creation of Magnetic Skyrmion Bubble Lattices by
  Ultrafast Laser in Ultrathin Films}}.
\newblock \bibinfo{journal}{Nano Letters}
  \bibinfo{year}{2018};\bibinfo{volume}{18}(\bibinfo{number}{11}):\bibinfo{pages}{7362--7371}.
\newblock \DOIprefix\doi{10.1021/acs.nanolett.8b03653}.
\bibitem[{Maccariello et~al.(2018)Maccariello, Legrand, Reyren, Garcia,
  Bouzehouane, Collin et~al.}]{Maccariello2018ElectricalTemperature}
\bibinfo{author}{Maccariello\xfnm[ D.]}, \bibinfo{author}{Legrand\xfnm[ W.]},
  \bibinfo{author}{Reyren\xfnm[ N.]}, \bibinfo{author}{Garcia\xfnm[ K.]},
  \bibinfo{author}{Bouzehouane\xfnm[ K.]}, \bibinfo{author}{Collin\xfnm[ S.]},
  et~al.
\newblock \bibinfo{title}{{Electrical detection of single magnetic skyrmions in
  metallic multilayers at room temperature}}.
\newblock \bibinfo{journal}{Nature Nanotechnology}
  \bibinfo{year}{2018};\bibinfo{volume}{13}(\bibinfo{number}{3}):\bibinfo{pages}{233--237}.
\newblock \DOIprefix\doi{10.1038/s41565-017-0044-4}.
\bibitem[{Heistracher et~al.(2018)Heistracher, Abert, Bruckner, Vogler and
  Suess}]{Heistracher2018GPU-AcceleratedAnnihilations}
\bibinfo{author}{Heistracher\xfnm[ P.]}, \bibinfo{author}{Abert\xfnm[ C.]},
  \bibinfo{author}{Bruckner\xfnm[ F.]}, \bibinfo{author}{Vogler\xfnm[ C.]},
  \bibinfo{author}{Suess\xfnm[ D.]}.
\newblock \bibinfo{title}{{GPU-Accelerated Atomistic Energy Barrier
  Calculations of Skyrmion Annihilations}}.
\newblock \bibinfo{journal}{IEEE Transactions on Magnetics}
  \bibinfo{year}{2018};\bibinfo{volume}{54}(\bibinfo{number}{11}).
\newblock \DOIprefix\doi{10.1109/TMAG.2018.2847446}.
\bibitem[{Ma et~al.(2015)Ma, Zhou, Braun and Lew}]{Ma2015}
\bibinfo{author}{Ma\xfnm[ F.]}, \bibinfo{author}{Zhou\xfnm[ Y.]},
  \bibinfo{author}{Braun\xfnm[ H.B.]}, \bibinfo{author}{Lew\xfnm[ W.S.]}.
\newblock \bibinfo{title}{{Skyrmion-Based Dynamic Magnonic Crystal}}.
\newblock \bibinfo{journal}{Nano Letters}
  \bibinfo{year}{2015};\bibinfo{volume}{15}(\bibinfo{number}{6}):\bibinfo{pages}{4029--4036}.
\newblock \DOIprefix\doi{10.1021/acs.nanolett.5b00996}.
\bibitem[{Fert et~al.(2017)Fert, Reyren and
  Cros}]{Fert_Magnetic_skyrmions_applications_2017}
\bibinfo{author}{Fert\xfnm[ A.]}, \bibinfo{author}{Reyren\xfnm[ N.]},
  \bibinfo{author}{Cros\xfnm[ V.]}.
\newblock \bibinfo{title}{{Magnetic skyrmions: advances in physics and
  potential applications}}.
\newblock \bibinfo{journal}{Nature Reviews Materials}
  \bibinfo{year}{2017};\bibinfo{volume}{2}(\bibinfo{number}{7}):\bibinfo{pages}{17031}.
\newblock \URLprefix \url{http://www.nature.com/articles/natrevmats201731}.
  \DOIprefix\doi{10.1038/natrevmats.2017.31}.
\bibitem[{Suess et~al.(2019)Suess, Vogler, Bruckner, Heistracher, Slanovc and
  Abert}]{Suess2019SpinMemory}
\bibinfo{author}{Suess\xfnm[ D.]}, \bibinfo{author}{Vogler\xfnm[ C.]},
  \bibinfo{author}{Bruckner\xfnm[ F.]}, \bibinfo{author}{Heistracher\xfnm[
  P.]}, \bibinfo{author}{Slanovc\xfnm[ F.]}, \bibinfo{author}{Abert\xfnm[ C.]}.
\newblock \bibinfo{title}{{Spin torque efficiency and analytic error rate
  estimates of skyrmion racetrack memory}}.
\newblock \bibinfo{journal}{Scientific Reports}
  \bibinfo{year}{2019};\bibinfo{volume}{9}(\bibinfo{number}{1}):\bibinfo{pages}{4827}.
\newblock \DOIprefix\doi{10.1038/s41598-019-41062-y}.
\bibitem[{Suess et~al.(2018)Suess, Vogler, Bruckner, Heistracher and
  Abert}]{Suess2018AMemory}
\bibinfo{author}{Suess\xfnm[ D.]}, \bibinfo{author}{Vogler\xfnm[ C.]},
  \bibinfo{author}{Bruckner\xfnm[ F.]}, \bibinfo{author}{Heistracher\xfnm[
  P.]}, \bibinfo{author}{Abert\xfnm[ C.]}.
\newblock \bibinfo{title}{{A repulsive skyrmion chain as a guiding track for a
  racetrack memory}}.
\newblock \bibinfo{journal}{AIP Advances}
  \bibinfo{year}{2018};\bibinfo{volume}{8}(\bibinfo{number}{11}):\bibinfo{pages}{115301}.
\newblock \DOIprefix\doi{10.1063/1.4993957}.
\bibitem[{Romming et~al.(2013)Romming, Hanneken, Menzel, Bickel, Wolter,
  Von~Bergmann et~al.}]{Romming636}
\bibinfo{author}{Romming\xfnm[ N.]}, \bibinfo{author}{Hanneken\xfnm[ C.]},
  \bibinfo{author}{Menzel\xfnm[ M.]}, \bibinfo{author}{Bickel\xfnm[ J.E.]},
  \bibinfo{author}{Wolter\xfnm[ B.]}, \bibinfo{author}{Von~Bergmann\xfnm[ K.]},
  et~al.
\newblock \bibinfo{title}{{Writing and deleting single magnetic skyrmions}}.
\newblock \bibinfo{journal}{Science}
  \bibinfo{year}{2013};\bibinfo{volume}{341}(\bibinfo{number}{6146}):\bibinfo{pages}{636--639}.
\newblock \DOIprefix\doi{10.1126/science.1240573}.
\bibitem[{B{\"{u}}ttner et~al.(2018)B{\"{u}}ttner, Lemesh and
  Beach}]{Buttner2018TheoryApplications}
\bibinfo{author}{B{\"{u}}ttner\xfnm[ F.]}, \bibinfo{author}{Lemesh\xfnm[ I.]},
  \bibinfo{author}{Beach\xfnm[ G.S.]}.
\newblock \bibinfo{title}{{Theory of isolated magnetic skyrmions: From
  fundamentals to room temperature applications}}.
\newblock \bibinfo{journal}{Scientific Reports}
  \bibinfo{year}{2018};\bibinfo{volume}{8}(\bibinfo{number}{1}):\bibinfo{pages}{4464}.
\newblock \URLprefix \url{http://www.nature.com/articles/s41598-018-22242-8}.
  \DOIprefix\doi{10.1038/s41598-018-22242-8}.
\bibitem[{Zelent et~al.(2017)Zelent, T{\'{o}}bik, Krawczyk, Guslienko and
  Mruczkiewicz}]{PSSR:PSSR201700259}
\bibinfo{author}{Zelent\xfnm[ M.]}, \bibinfo{author}{T{\'{o}}bik\xfnm[ J.]},
  \bibinfo{author}{Krawczyk\xfnm[ M.]}, \bibinfo{author}{Guslienko\xfnm[
  K.Y.]}, \bibinfo{author}{Mruczkiewicz\xfnm[ M.]}.
\newblock \bibinfo{title}{{Bi-stability of magnetic skyrmions in ultrathin
  multilayer nNanodots induced by magnetostatic interaction}}.
\newblock \bibinfo{journal}{Physica Status Solidi - Rapid Research Letters}
  \bibinfo{year}{2017};\bibinfo{volume}{11}(\bibinfo{number}{10}):\bibinfo{pages}{1700259}.
\newblock \DOIprefix\doi{10.1002/pssr.201700259}.
\bibitem[{Rohart and Thiaville(2013)}]{Rohart2013SkyrmionInteraction}
\bibinfo{author}{Rohart\xfnm[ S.]}, \bibinfo{author}{Thiaville\xfnm[ A.]}.
\newblock \bibinfo{title}{{Skyrmion confinement in ultrathin film
  nanostructures in the presence of Dzyaloshinskii-Moriya interaction}}.
\newblock \bibinfo{journal}{Physical Review B}
  \bibinfo{year}{2013};\bibinfo{volume}{88}(\bibinfo{number}{18}):\bibinfo{pages}{184422}.
\newblock \URLprefix \url{https://link.aps.org/doi/10.1103/PhysRevB.88.184422}.
  \DOIprefix\doi{10.1103/PhysRevB.88.184422}.
\bibitem[{Zeissler et~al.(2017)Zeissler, Mruczkiewicz, Finizio, Raabe, Shepley,
  Sadovnikov et~al.}]{Zeissler2017PinningStacks}
\bibinfo{author}{Zeissler\xfnm[ K.]}, \bibinfo{author}{Mruczkiewicz\xfnm[ M.]},
  \bibinfo{author}{Finizio\xfnm[ S.]}, \bibinfo{author}{Raabe\xfnm[ J.]},
  \bibinfo{author}{Shepley\xfnm[ P.M.]}, \bibinfo{author}{Sadovnikov\xfnm[
  A.V.]}, et~al.
\newblock \bibinfo{title}{{Pinning and hysteresis in the field dependent
  diameter evolution of skyrmions in Pt/Co/Ir superlattice stacks}}.
\newblock \bibinfo{journal}{Scientific Reports}
  \bibinfo{year}{2017};\bibinfo{volume}{7}(\bibinfo{number}{1}):\bibinfo{pages}{1706.01065}.
\newblock \URLprefix \url{http://arxiv.org/abs/1706.01065}.
  \DOIprefix\doi{10.1038/s41598-017-15262-3}.
\bibitem[{Karakas et~al.(2018)Karakas, Gokce, Habiboglu, Arpaci, Ozbozduman,
  Cinar et~al.}]{Karakas2018ObservationAnisotropy}
\bibinfo{author}{Karakas\xfnm[ V.]}, \bibinfo{author}{Gokce\xfnm[ A.]},
  \bibinfo{author}{Habiboglu\xfnm[ A.T.]}, \bibinfo{author}{Arpaci\xfnm[ S.]},
  \bibinfo{author}{Ozbozduman\xfnm[ K.]}, \bibinfo{author}{Cinar\xfnm[ I.]},
  et~al.
\newblock \bibinfo{title}{{Observation of Magnetic Radial Vortex Nucleation in
  a Multilayer Stack with Tunable Anisotropy}}.
\newblock \bibinfo{journal}{Scientific Reports}
  \bibinfo{year}{2018};\bibinfo{volume}{8}(\bibinfo{number}{1}).
\newblock \DOIprefix\doi{10.1038/s41598-018-25392-x}.
\bibitem[{Boulle et~al.(2016)Boulle, Vogel, Yang, Pizzini, de~Souza~Chaves,
  Locatelli et~al.}]{boulle2016room}
\bibinfo{author}{Boulle\xfnm[ O.]}, \bibinfo{author}{Vogel\xfnm[ J.]},
  \bibinfo{author}{Yang\xfnm[ H.]}, \bibinfo{author}{Pizzini\xfnm[ S.]},
  \bibinfo{author}{de~Souza~Chaves\xfnm[ D.]}, \bibinfo{author}{Locatelli\xfnm[
  A.]}, et~al.
\newblock \bibinfo{title}{{Room-temperature chiral magnetic skyrmions in
  ultrathin magnetic nanostructures}}.
\newblock \bibinfo{journal}{Nature Nanotechnology}
  \bibinfo{year}{2016};\bibinfo{volume}{11}(\bibinfo{number}{5}):\bibinfo{pages}{449--454}.
\newblock \URLprefix \url{http://dx.doi.org/10.1038/nnano.2015.315}.
  \DOIprefix\doi{10.1038/nnano.2015.315}.
\bibitem[{Saha et~al.(2019)Saha, Zelent, Finizio, Mruczkiewicz, Tacchi, Suszka
  et~al.}]{Saha2019FormationAnisotropy}
\bibinfo{author}{Saha\xfnm[ S.]}, \bibinfo{author}{Zelent\xfnm[ M.]},
  \bibinfo{author}{Finizio\xfnm[ S.]}, \bibinfo{author}{Mruczkiewicz\xfnm[
  M.]}, \bibinfo{author}{Tacchi\xfnm[ S.]}, \bibinfo{author}{Suszka\xfnm[
  A.K.]}, et~al.
\newblock \bibinfo{title}{{Formation of N{\'{e}}el-type skyrmions in an antidot
  lattice with perpendicular magnetic anisotropy}}.
\newblock \bibinfo{journal}{Physical Review B}
  \bibinfo{year}{2019};\bibinfo{volume}{100}(\bibinfo{number}{14}):\bibinfo{pages}{144435}.
\newblock \URLprefix
  \url{https://link.aps.org/doi/10.1103/PhysRevB.100.144435}.
  \DOIprefix\doi{10.1103/PhysRevB.100.144435}.
\bibitem[{Casiraghi et~al.(2019)Casiraghi, Corte-Le{\'{o}}n, Vafaee,
  Garcia-Sanchez, Durin, Pasquale et~al.}]{Casiraghi2019IndividualGradients}
\bibinfo{author}{Casiraghi\xfnm[ A.]}, \bibinfo{author}{Corte-Le{\'{o}}n\xfnm[
  H.]}, \bibinfo{author}{Vafaee\xfnm[ M.]},
  \bibinfo{author}{Garcia-Sanchez\xfnm[ F.]}, \bibinfo{author}{Durin\xfnm[
  G.]}, \bibinfo{author}{Pasquale\xfnm[ M.]}, et~al.
\newblock \bibinfo{title}{{Individual skyrmion manipulation by local magnetic
  field gradients}}.
\newblock \bibinfo{journal}{Communications Physics}
  \bibinfo{year}{2019};\bibinfo{volume}{2}(\bibinfo{number}{1}).
\newblock \DOIprefix\doi{10.1038/s42005-019-0242-5}.
\bibitem[{Qin et~al.(2018)Qin, Jin, Xie, Li, Wang, Cao
  et~al.}]{Qin2018Size-tunableMultilayers}
\bibinfo{author}{Qin\xfnm[ Z.]}, \bibinfo{author}{Jin\xfnm[ C.]},
  \bibinfo{author}{Xie\xfnm[ H.]}, \bibinfo{author}{Li\xfnm[ X.]},
  \bibinfo{author}{Wang\xfnm[ Y.]}, \bibinfo{author}{Cao\xfnm[ J.]}, et~al.
\newblock \bibinfo{title}{{Size-tunable skyrmion bubbles in Ta/CoFeB/MgO
  multilayers}}.
\newblock \bibinfo{journal}{Journal of Physics D: Applied Physics}
  \bibinfo{year}{2018};\bibinfo{volume}{51}(\bibinfo{number}{42}).
\newblock \DOIprefix\doi{10.1088/1361-6463/aadd59}.
\bibitem[{Temiryazev et~al.(2018)Temiryazev, Temiryazeva, Zdoroveyshchev,
  Vikhrova, Dorokhin, Demina et~al.}]{Temiryazev2018FormationMicroscope}
\bibinfo{author}{Temiryazev\xfnm[ A.G.]}, \bibinfo{author}{Temiryazeva\xfnm[
  M.P.]}, \bibinfo{author}{Zdoroveyshchev\xfnm[ A.V.]},
  \bibinfo{author}{Vikhrova\xfnm[ O.V.]}, \bibinfo{author}{Dorokhin\xfnm[
  M.V.]}, \bibinfo{author}{Demina\xfnm[ P.B.]}, et~al.
\newblock \bibinfo{title}{{Formation of a Domain Structure in Multilayer CoPt
  Films by Magnetic Probe of an Atomic Force Microscope}}.
\newblock \bibinfo{journal}{Physics of the Solid State}
  \bibinfo{year}{2018};\bibinfo{volume}{60}(\bibinfo{number}{11}):\bibinfo{pages}{2200--2206}.
\newblock \DOIprefix\doi{10.1134/S1063783418110318}.
\bibitem[{Zhang et~al.(2018)Zhang, Zhang, Zhang, Barton, Neu, Zhao
  et~al.}]{Zhang2018DirectField}
\bibinfo{author}{Zhang\xfnm[ S.]}, \bibinfo{author}{Zhang\xfnm[ J.]},
  \bibinfo{author}{Zhang\xfnm[ Q.]}, \bibinfo{author}{Barton\xfnm[ C.]},
  \bibinfo{author}{Neu\xfnm[ V.]}, \bibinfo{author}{Zhao\xfnm[ Y.]}, et~al.
\newblock \bibinfo{title}{{Direct writing of room temperature and zero field
  skyrmion lattices by a scanning local magnetic field}}.
\newblock \bibinfo{journal}{Applied Physics Letters}
  \bibinfo{year}{2018};\bibinfo{volume}{112}(\bibinfo{number}{13}).
\newblock \DOIprefix\doi{10.1063/1.5021172}.
\bibitem[{Mironov et~al.(2019)Mironov, Gorev, Ermolaeva, Gusev and
  Petrov}]{Mironov2019ImpactAnisotropy}
\bibinfo{author}{Mironov\xfnm[ V.L.]}, \bibinfo{author}{Gorev\xfnm[ R.V.]},
  \bibinfo{author}{Ermolaeva\xfnm[ O.L.]}, \bibinfo{author}{Gusev\xfnm[ N.S.]},
  \bibinfo{author}{Petrov\xfnm[ Y.V.]}.
\newblock \bibinfo{title}{{Impact of the Field of a Magnetic Force Microscope
  Probe on the Skyrmion State in a Modified Co/Pt Film with Perpendicular
  Anisotropy}}.
\newblock \bibinfo{journal}{Physics of the Solid State}
  \bibinfo{year}{2019};\bibinfo{volume}{61}(\bibinfo{number}{9}):\bibinfo{pages}{1594--1598}.
\newblock \DOIprefix\doi{10.1134/S1063783419090154}.
\bibitem[{Berganza et~al.(2018)Berganza, Jaafar, Goiriena-Goikoetxea,
  Pablo-Navarro, Garcia-Arribas, Gusliyenko
  et~al.}]{Berganza2018ObservationNanodots}
\bibinfo{author}{Berganza\xfnm[ E.]}, \bibinfo{author}{Jaafar\xfnm[ M.]},
  \bibinfo{author}{Goiriena-Goikoetxea\xfnm[ M.]},
  \bibinfo{author}{Pablo-Navarro\xfnm[ J.]},
  \bibinfo{author}{Garcia-Arribas\xfnm[ A.]}, \bibinfo{author}{Gusliyenko\xfnm[
  K.]}, et~al.
\newblock \bibinfo{title}{{Observation of hedgehog skyrmions in sub-100 nm soft
  magnetic nanodots}} \bibinfo{year}{2018};\URLprefix
  \url{http://arxiv.org/abs/1803.08768}.
\bibitem[{Vansteenkiste and Van~de Wiele(2011)}]{MuMax2011_main}
\bibinfo{author}{Vansteenkiste\xfnm[ A.]}, \bibinfo{author}{Van~de Wiele\xfnm[
  B.]}.
\newblock \bibinfo{title}{{MuMax: A new high-performance micromagnetic
  simulation tool}}.
\newblock \bibinfo{journal}{Journal of Magnetism and Magnetic Materials}
  \bibinfo{year}{2011};\bibinfo{volume}{323}(\bibinfo{number}{21}):\bibinfo{pages}{2585--2591}.
\newblock \DOIprefix\doi{10.1016/j.jmmm.2011.05.037}.
\bibitem[{Vansteenkiste et~al.(2014)Vansteenkiste, Leliaert, Dvornik, Helsen,
  Garcia-Sanchez and Van~Waeyenberge}]{mumax_2014}
\bibinfo{author}{Vansteenkiste\xfnm[ A.]}, \bibinfo{author}{Leliaert\xfnm[
  J.]}, \bibinfo{author}{Dvornik\xfnm[ M.]}, \bibinfo{author}{Helsen\xfnm[
  M.]}, \bibinfo{author}{Garcia-Sanchez\xfnm[ F.]},
  \bibinfo{author}{Van~Waeyenberge\xfnm[ B.]}.
\newblock \bibinfo{title}{{The design and verification of MuMax3}}.
\newblock \bibinfo{journal}{AIP Advances}
  \bibinfo{year}{2014};\bibinfo{volume}{4}(\bibinfo{number}{10}):\bibinfo{pages}{107133}.
\newblock \DOIprefix\doi{10.1063/1.4899186}.
\bibitem[{Leliaert et~al.(2014)Leliaert, Van~de Wiele, Vansteenkiste, Laurson,
  Durin, Dupr{\'{e}} et~al.}]{Leliaert2014}
\bibinfo{author}{Leliaert\xfnm[ J.]}, \bibinfo{author}{Van~de Wiele\xfnm[ B.]},
  \bibinfo{author}{Vansteenkiste\xfnm[ A.]}, \bibinfo{author}{Laurson\xfnm[
  L.]}, \bibinfo{author}{Durin\xfnm[ G.]}, \bibinfo{author}{Dupr{\'{e}}\xfnm[
  L.]}, et~al.
\newblock \bibinfo{title}{{Current-driven domain wall mobility in
  polycrystalline Permalloy nanowires: A numerical study}}.
\newblock \bibinfo{journal}{Journal of Applied Physics}
  \bibinfo{year}{2014};\bibinfo{volume}{115}(\bibinfo{number}{23}):\bibinfo{pages}{233903}.
\newblock \URLprefix \url{http://aip.scitation.org/doi/10.1063/1.4883297}.
  \DOIprefix\doi{10.1063/1.4883297}.
\bibitem[{Bautin et~al.(2017)Bautin, Seferyan, Nesmeyanov and
  Usov}]{Bautin2017MagneticNanoparticles}
\bibinfo{author}{Bautin\xfnm[ V.A.]}, \bibinfo{author}{Seferyan\xfnm[ A.G.]},
  \bibinfo{author}{Nesmeyanov\xfnm[ M.S.]}, \bibinfo{author}{Usov\xfnm[ N.A.]}.
\newblock \bibinfo{title}{{Magnetic properties of polycrystalline cobalt
  nanoparticles}}.
\newblock \bibinfo{journal}{AIP Advances}
  \bibinfo{year}{2017};\bibinfo{volume}{7}(\bibinfo{number}{4}):\bibinfo{pages}{045103}.
\newblock \DOIprefix\doi{10.1063/1.4979889}.
\bibitem[{Di et~al.(2015)Di, Zhang, Lim, Ng, Kuok, Yu et~al.}]{bls1}
\bibinfo{author}{Di\xfnm[ K.]}, \bibinfo{author}{Zhang\xfnm[ V.L.]},
  \bibinfo{author}{Lim\xfnm[ H.S.]}, \bibinfo{author}{Ng\xfnm[ S.C.]},
  \bibinfo{author}{Kuok\xfnm[ M.H.]}, \bibinfo{author}{Yu\xfnm[ J.]}, et~al.
\newblock \bibinfo{title}{Direct observation of the dzyaloshinskii-moriya
  interaction in a pt/co/ni film}.
\newblock \bibinfo{journal}{Phys Rev Lett}
  \bibinfo{year}{2015};\bibinfo{volume}{114}:\bibinfo{pages}{047201}.
\bibitem[{Samardak et~al.(2018)Samardak, Kolesnikov, Stebliy, Chebotkevich,
  Sadovnikov, Nikitov et~al.}]{bls2}
\bibinfo{author}{Samardak\xfnm[ A.]}, \bibinfo{author}{Kolesnikov\xfnm[ A.]},
  \bibinfo{author}{Stebliy\xfnm[ M.]}, \bibinfo{author}{Chebotkevich\xfnm[
  L.]}, \bibinfo{author}{Sadovnikov\xfnm[ A.]}, \bibinfo{author}{Nikitov\xfnm[
  S.]}, et~al.
\newblock \bibinfo{title}{Enhanced interfacial dzyaloshinskii-moriya
  interaction and isolated skyrmions in the inversion-symmetry-broken
  ru/co/w/ru films}.
\newblock \bibinfo{journal}{Applied Physics Letters}
  \bibinfo{year}{2018};\bibinfo{volume}{112}(\bibinfo{number}{19}):\bibinfo{pages}{192406}.
\bibitem[{Sadovnikov et~al.(2019)Sadovnikov, Beginin, Sheshukova, Sharaevskii,
  Stognij, Novitski et~al.}]{bls3}
\bibinfo{author}{Sadovnikov\xfnm[ A.V.]}, \bibinfo{author}{Beginin\xfnm[
  E.N.]}, \bibinfo{author}{Sheshukova\xfnm[ S.E.]},
  \bibinfo{author}{Sharaevskii\xfnm[ Y.P.]}, \bibinfo{author}{Stognij\xfnm[
  A.I.]}, \bibinfo{author}{Novitski\xfnm[ N.N.]}, et~al.
\newblock \bibinfo{title}{Route toward semiconductor magnonics: Light-induced
  spin-wave nonreciprocity in a yig/gaas structure}.
\newblock \bibinfo{journal}{Phys Rev B}
  \bibinfo{year}{2019};\bibinfo{volume}{99}:\bibinfo{pages}{054424}.
\bibitem[{Zeissler et~al.(2018)Zeissler, Finizio, Shahbazi, Massey, Ma’Mari,
  Bracher et~al.}]{Zeissler2018DiscreteNanodiscs}
\bibinfo{author}{Zeissler\xfnm[ K.]}, \bibinfo{author}{Finizio\xfnm[ S.]},
  \bibinfo{author}{Shahbazi\xfnm[ K.]}, \bibinfo{author}{Massey\xfnm[ J.]},
  \bibinfo{author}{Ma’Mari\xfnm[ F.A.]}, \bibinfo{author}{Bracher\xfnm[
  D.M.]}, et~al.
\newblock \bibinfo{title}{{Discrete Hall resistivity contribution from
  N{\'{e}}el skyrmions in multilayer nanodiscs}}.
\newblock \bibinfo{journal}{Nature Nanotechnology}
  \bibinfo{year}{2018};\bibinfo{volume}{13}(\bibinfo{number}{12}):\bibinfo{pages}{1161--1166}.
\newblock \DOIprefix\doi{10.1038/s41565-018-0268-y}.
\bibitem[{Bryan et~al.(2005)Bryan, Atkinson and
  Cowburn}]{Bryan2005EdgeNanostructures}
\bibinfo{author}{Bryan\xfnm[ M.T.]}, \bibinfo{author}{Atkinson\xfnm[ D.]},
  \bibinfo{author}{Cowburn\xfnm[ R.P.]}.
\newblock \bibinfo{title}{{Edge roughness and coercivity in magnetic
  nanostructures}}.
\newblock \bibinfo{journal}{Journal of Physics: Conference Series}
  \bibinfo{year}{2005};\bibinfo{volume}{17}(\bibinfo{number}{1}):\bibinfo{pages}{40--44}.
\newblock \URLprefix
  \url{http://stacks.iop.org/1742-6596/17/i=1/a=006?key=crossref.37de17e183fc02cd3c09f275a98f7ed4}.
  \DOIprefix\doi{10.1088/1742-6596/17/1/006}.
\bibitem[{P{\"{u}}tter et~al.(2009)P{\"{u}}tter, Mikuszeit, Vedmedenko and
  Oepen}]{Putter2009TheElements}
\bibinfo{author}{P{\"{u}}tter\xfnm[ S.]}, \bibinfo{author}{Mikuszeit\xfnm[
  N.]}, \bibinfo{author}{Vedmedenko\xfnm[ E.Y.]}, \bibinfo{author}{Oepen\xfnm[
  H.P.]}.
\newblock \bibinfo{title}{{The effect of tilted edges on the shape anisotropy
  and stray field coupling of uniformly magnetized rectangular elements}}.
\newblock \bibinfo{journal}{Journal of Applied Physics}
  \bibinfo{year}{2009};\bibinfo{volume}{106}(\bibinfo{number}{4}).
\newblock \DOIprefix\doi{10.1063/1.3169781}.
\bibitem[{Maranville et~al.(2007)Maranville, McMichael and
  Abraham}]{Maranville2007VariationStripes}
\bibinfo{author}{Maranville\xfnm[ B.B.]}, \bibinfo{author}{McMichael\xfnm[
  R.D.]}, \bibinfo{author}{Abraham\xfnm[ D.W.]}.
\newblock \bibinfo{title}{{Variation of thin film edge magnetic properties with
  patterning process conditions in Ni80Fe20 stripes}}.
\newblock \bibinfo{journal}{Applied Physics Letters}
  \bibinfo{year}{2007};\bibinfo{volume}{90}(\bibinfo{number}{23}):\bibinfo{pages}{20--23}.
\newblock \DOIprefix\doi{10.1063/1.2746406}.
\bibitem[{Zhu and McMichael(2010)}]{Zhu2010ModificationEdges}
\bibinfo{author}{Zhu\xfnm[ M.]}, \bibinfo{author}{McMichael\xfnm[ R.D.]}.
\newblock \bibinfo{title}{{Modification of edge mode dynamics by oxidation in
  Ni80 Fe 20 thin film edges}}.
\newblock \bibinfo{journal}{Journal of Applied Physics}
  \bibinfo{year}{2010};\bibinfo{volume}{107}(\bibinfo{number}{10}).
\newblock \DOIprefix\doi{10.1063/1.3393966}.
\bibitem[{Guo et~al.(2013)Guo, Belova and
  McMichael}]{Guo2013SpectroscopyNanodisks}
\bibinfo{author}{Guo\xfnm[ F.]}, \bibinfo{author}{Belova\xfnm[ L.M.]},
  \bibinfo{author}{McMichael\xfnm[ R.D.]}.
\newblock \bibinfo{title}{{Spectroscopy and Imaging of Edge Modes in Permalloy
  Nanodisks}}.
\newblock \bibinfo{journal}{Physical Review Letters}
  \bibinfo{year}{2013};\bibinfo{volume}{110}(\bibinfo{number}{1}):\bibinfo{pages}{017601}.
\newblock \URLprefix
  \url{https://link.aps.org/doi/10.1103/PhysRevLett.110.017601}.
  \DOIprefix\doi{10.1103/PhysRevLett.110.017601}.
\bibitem[{Metaxas et~al.(2007)Metaxas, Jamet, Mougin, Cormier, Ferre, Baltz
  et~al.}]{Metaxas2007}
\bibinfo{author}{Metaxas\xfnm[ P.J.]}, \bibinfo{author}{Jamet\xfnm[ J.P.]},
  \bibinfo{author}{Mougin\xfnm[ A.]}, \bibinfo{author}{Cormier\xfnm[ M.]},
  \bibinfo{author}{Ferre\xfnm[ J.]}, \bibinfo{author}{Baltz\xfnm[ V.]}, et~al.
\newblock \bibinfo{title}{{Creep and Flow Regimes of Magnetic Domain-Wall
  Motion in Ultrathin Pt/Co/Pt Films with Perpendicular Anisotropy}}.
\newblock \bibinfo{journal}{Phys Rev Lett}
  \bibinfo{year}{2007};\bibinfo{volume}{99}:\bibinfo{pages}{217208}.
\bibitem[{Mironov et~al.(2009)Mironov, Gribkov, Vdovichev, Gusev, Fraerman,
  Ermolaeva et~al.}]{Mironov2009MagneticAnisotropy}
\bibinfo{author}{Mironov\xfnm[ V.L.]}, \bibinfo{author}{Gribkov\xfnm[ B.A.]},
  \bibinfo{author}{Vdovichev\xfnm[ S.N.]}, \bibinfo{author}{Gusev\xfnm[ S.A.]},
  \bibinfo{author}{Fraerman\xfnm[ A.A.]}, \bibinfo{author}{Ermolaeva\xfnm[
  O.L.]}, et~al.
\newblock \bibinfo{title}{{Magnetic force microscope tip-induced
  remagnetization of CoPt nanodisks with perpendicular anisotropy}}.
\newblock \bibinfo{journal}{Journal of Applied Physics}
  \bibinfo{year}{2009};\bibinfo{volume}{106}(\bibinfo{number}{5}).
\newblock \DOIprefix\doi{10.1063/1.3202354}.
\bibitem[{Gorchon et~al.(2014)Gorchon, Bustingorry, Ferr{\'{e}}, Jeudy, Kolton
  and Giamarchi}]{Gorchon2014Pinning-dependentFilm}
\bibinfo{author}{Gorchon\xfnm[ J.]}, \bibinfo{author}{Bustingorry\xfnm[ S.]},
  \bibinfo{author}{Ferr{\'{e}}\xfnm[ J.]}, \bibinfo{author}{Jeudy\xfnm[ V.]},
  \bibinfo{author}{Kolton\xfnm[ A.B.]}, \bibinfo{author}{Giamarchi\xfnm[ T.]}.
\newblock \bibinfo{title}{{Pinning-dependent field-driven domain wall dynamics
  and thermal scaling in an ultrathin Pt/Co/Pt magnetic film}}.
\newblock \bibinfo{journal}{Physical Review Letters}
  \bibinfo{year}{2014};\bibinfo{volume}{113}(\bibinfo{number}{2}):\bibinfo{pages}{027205}.
\newblock \DOIprefix\doi{10.1103/PhysRevLett.113.027205}.
\bibitem[{Vettiger et~al.(2002)Vettiger, Cross, Despont, Drechsler, Durig,
  Gotsmann et~al.}]{Vettiger2002TheStorage}
\bibinfo{author}{Vettiger\xfnm[ P.]}, \bibinfo{author}{Cross\xfnm[ G.]},
  \bibinfo{author}{Despont\xfnm[ M.]}, \bibinfo{author}{Drechsler\xfnm[ U.]},
  \bibinfo{author}{Durig\xfnm[ U.]}, \bibinfo{author}{Gotsmann\xfnm[ B.]},
  et~al.
\newblock \bibinfo{title}{{The "millipede" - nanotechnology entering data
  storage}}.
\newblock \bibinfo{journal}{IEEE Transactions on Nanotechnology}
  \bibinfo{year}{2002};\bibinfo{volume}{1}(\bibinfo{number}{1}):\bibinfo{pages}{39--55}.
\newblock \URLprefix \url{https://ieeexplore.ieee.org/document/1005425/}.
  \DOIprefix\doi{10.1109/TNANO.2002.1005425}.

\end{thebibliography}
\end{document}